\documentstyle[prl,twocolumn,aps,cite]{revtex}
\begin{document}
\draft

 \newcommand{\mytitle}[1]{
 \twocolumn[\hsize\textwidth\columnwidth\hsize
 \csname@twocolumnfalse\endcsname #1 \vspace{1mm}]}

\mytitle{
\title{Coherence and Partial Coherence in Interacting Electron Systems}

\author{J\"urgen K\"onig$^{1,2}$ and Yuval Gefen$^{3}$}
\address{
$^1$Department of Physics, The University of Texas at Austin, Austin, Texas
78712, USA\\
$^2$Department of Physics, Indiana University, Bloomington,
Indiana 47405, USA\\
$^3$Department of Condensed Matter Physics, The Weizmann Institute of Science, 
76100 Rehovot, Israel}

\date{\today}  

\maketitle

\begin{abstract}
We study coherence of electron transport through interacting quantum dots and
discuss the relation of the coherent part to the flux-sensitive conductance
for three different types of Aharonov-Bohm interferometers.
Contributions to transport in first and second order in the intrinsic line 
width of the dot levels are addressed in detail.
We predict an asymmetry of the interference signal around resonance peaks
as a consequence of incoherence associated with spin-flip processes.
Furthermore, we show by strict calculation that first-order contributions can 
be partially or even fully coherent.
This contrasts with the sequential-tunneling picture which describes 
first-order transport as a sequence of incoherent processes.
\end{abstract}
}

{\it Introduction.}
The study of transport through quantum dots (QDs) revealed interesting 
phenomena like resonant tunneling, Coulomb blockade, and the Kondo effect.
The measurement of the current, however, provides no information whether this 
transport occurs coherently or incoherently.
To approach this question the QD can be embedded in an Aharonov-Bohm (AB) 
geometry
\cite{Yacobi95,Yeyati95,Hackenbroich96,Bruder96,Oreg97,Gerland00,Schuster97}. 
A magnetic-flux sensitivity of the total current has been observed 
\cite{Yacobi95,Schuster97}.
Depending on the setup, the phase of the oscillations showed either a jump as 
a function of the gate voltage or a continuous phase shift.

The transmission probability $T^{\rm dot}_\sigma(\omega)$ through a QD for
incoming electrons with energy $\omega$ and spin $\sigma$ is defined by its 
relation to the linear conductance,
\begin{equation}
\label{conductance-transmission}
   {\partial I\over \partial V}\bigg|_{V=0} = 
	- {e^2\over h}\sum_\sigma \int
	d \omega \, T_\sigma(\omega) f'(\omega) \, .
\end{equation}
In the absence of electron-electron interaction, transport can be described 
within a scattering approach \cite{Landauer70,Buettiker85} with a 
transmission amplitude 
$t_\sigma^{\rm dot}(\omega) \propto G^{\rm ret}_{\sigma,\rm LR} (\omega)$
and $T^{\rm dot}_\sigma(\omega) = |t^{\rm dot}_\sigma(\omega)|^2$. 
The Green's function $G^{\rm ret}_{\sigma,\rm LR} (\omega)$ involves Fermi 
operators of the left and the right electron reservoirs.
It is, then, easy to show that transport through the QD is
fully coherent.
At low temperature it is possible to tune the transmission amplitude of a
reference arm {\it and} the AB flux such that the total transmission is zero, 
corresponding to a fully destructive interference. 
In the presence of electron-electron interaction, though, this approach fails.
The transmission probability $T^{\rm dot}_\sigma(\omega)$ 
can no longer be obtained from $t^{\rm dot}_{\sigma}(\omega)$ introduced above
but has to be determined using Green's function techniques for interacting 
systems \cite{Meir92,Koenig96,Koenig99}.
Thus, the question of whether and how the coherent part of the transport 
through an {\it interacting} QD can be identified is nontrivial.
It will be addressed in this Letter.

First we use intuitive arguments to distinguish coherent from incoherent 
cotunneling through a noninteracting and an interacting QD.
Then, for a quantitative analysis, we develop general expressions for the 
flux-sensitive transmission through an interferometer containing either one 
or two QDs.
We derive the intuitively obvious result that the coherence of cotunneling may 
be spoiled by spin-flip processes.
They give rise to an incoherence-induced asymmetry of the amplitude of the 
interference signal.
We propose a symmetric AB interferometer using two QDs to show that
first-order transport can be partially or even fully coherent, in contrast to
the description of first-order transport within the language of incoherent
sequential tunneling.

We consider a single-level QD with level energy $\epsilon$, measured from 
the Fermi energy of the leads.
The Hamiltonian $H=H_L+H_R+H_D+H_T$ contains 
$H_r = \sum_{k \sigma} \epsilon_{kr} a^\dagger_{k\sigma r}a_{k\sigma r}$
for the left and right lead, $r=L/R$.
The isolated dot is described by 
$H_D = \epsilon \sum_\sigma n_\sigma + U n_\uparrow n_\downarrow$, where 
$n_\sigma = c^\dagger_\sigma c_\sigma$, and $H_T = \sum_{k \sigma r} (t_r 
a^\dagger_{k\sigma r} c_\sigma + {\rm H.c.})$ models tunneling between dot 
and leads (we neglect the energy dependence of the tunnel matrix elements 
$t_{L/R}$).
Due to tunneling the dot level acquires a finite line width 
$\Gamma=\Gamma_L+\Gamma_R$ with $\Gamma_{L/R}=2\pi|t_{L/R}|^2N_{L/R}$ where 
$N_{L/R}$ is the density of states in the leads.
The electron-electron interaction is accounted for by the charging energy 
$U=2E_C$ for double occupancy.
To keep the discussion simple we choose $U=0$ for the noninteracting case and 
$U=\infty$ for an interacting QD.

{\it Transport through noninteracting and interacting QDs.}
In the absence of interaction we can use the scattering formalism with
the transmission amplitude
\begin{equation}
\label{amplitude}
   t_\sigma^{\rm dot}(\omega) = i \sqrt{\Gamma_L \Gamma_R} \, 
	G^{\rm ret}_\sigma (\omega) \, ,
\end{equation}
where the dot Green's function 
$G^{\rm ret}_\sigma (\omega)=1/(\omega-\epsilon + i\Gamma/2)$ is the 
Fourier transform of 
$-i\Theta (t) \langle \left\{ c_\sigma (t),c^\dagger_\sigma (0) \right\} 
\rangle$.
The transmission probability $T^{\rm dot}_\sigma(\omega)= 
|t^{\rm dot}_\sigma(\omega)|^2 = \Gamma_L \Gamma_R / 
[(\omega - \epsilon)^2 + (\Gamma/2)^2]$ reflects resonant tunneling, which 
is fully coherent to all orders in $\Gamma$.

We identify a contribution to the transport through a QD as coherent if by 
{\it adding} a reference trajectory {\it fully destructive} interference can 
be achieved.
Interaction of the dot electrons with an external bath (e.g. phonons) destroys 
coherence since interference with a reference trajectory is no longer possible:
the transmitted electron has changed its state or, equivalently \cite{Stern90},
a trace in the environment is left.
Coherence can be also lost in {\it interacting} QDs by flipping both the 
spin of the transmitted electron and the QD.

Away from resonance, $|\epsilon| \gg k_BT, \Gamma$, transport is dominated by
cotunneling \cite{Averin89,renormalization}.
There are three different types of cotunneling processes (for $U=\infty$):\\
(i) an electron enters the QD, leading to a virtual occupancy, and then 
leaves it to the other side.\\
(ii) an electron leaves the QD, and an electron with the same spin enters.\\
(iii) an electron leaves the QD, and an electron with opposite spin enters.
\\
Process (iii) is elastic in the sense that the energy of the QD has not 
changed.
It is incoherent, though, since the spin in the QD has flipped.

The transmission, defined by Eq.~(\ref{conductance-transmission}), 
through a single-level QD can be obtained \cite{Meir92,Koenig96,Koenig99} from
\begin{equation}
   T^{\rm dot}_\sigma(\omega) = -{2\Gamma_L \Gamma_R \over \Gamma} \,{\rm Im}\,
	G^{\rm ret}_\sigma(\omega) \, .
\end{equation}
For cotunneling, the transmission probabilities of electrons with energy 
$\omega$ near the Fermi level of the leads can also be obtained by 
calculating the transition rate in second-order perturbation theory and 
multiplying it with the probabilities $P_\chi$ to find the system in the 
corresponding initial state $\chi$.
For an incoming electron with spin up the transmission probabilities are
$P_\chi \Gamma_L \Gamma_R {\rm Re}[1/(\omega-\epsilon+i0^+)^2]$ with 
$\chi=0,\uparrow,\downarrow$ for case (i), (ii), and (iii), respectively.
Since $P_0 +P_\uparrow +P_\downarrow =1$ and $P_0+P_\sigma =1/[1+f(\epsilon)]$
in equilibrium, where $f(\epsilon)$ is the Fermi function, we find 
$T^{\rm dot}_\sigma(\omega) = T^{\rm dot, coh}_\sigma(\omega) + 
T^{\rm dot, incoh}_\sigma(\omega)$ with \cite{comment_reg}
\begin{eqnarray}
\label{cotunneling total}
   T^{\rm dot}_\sigma(\omega) &=& {\rm Re} \, {\Gamma_L \Gamma_R \over 
	(\omega - \epsilon + i0^+)^2} \, ,
\\
\label{cotunneling coherent}
   T^{\rm dot, coh}_\sigma(\omega) &=& 
	{T^{\rm dot}_\sigma(\omega)\over 1+f(\epsilon)} \, .
\end{eqnarray}

We now show that Eq.~(\ref{amplitude}) is {\it not} a good definition for a 
transmission amplitude for {\it interacting} QDs.
From $t_\sigma^{\rm dot}(\omega)= i (P_0 +P_\uparrow) \sqrt{\Gamma_L \Gamma_R}/
(\omega-\epsilon + i0^+)$ we get
$|t_\sigma^{\rm dot}(\omega)|^2 = T^{\rm dot}_\sigma(\omega)/[1+f(\epsilon)]^2$
which not only does not yield the total transmission through the dot but also 
differs from the coherent part of the transmission as well: there is no direct 
physical meaning of the expression $|t_\sigma^{\rm dot}(\omega)|^2$.

A generalization of the scattering approach has been proposed \cite{Imry99} 
which is compatible with the physical quantities expressed by 
Eqs.~(\ref{cotunneling total}) and (\ref{cotunneling coherent}).
While this generalization is physically motivated, it gives no recipe how to
calculate the transmission amplitudes explicitly in a given order in $\Gamma$.

For $U=0$, there are three more cotunneling processes.
They involve double occupancy as an intermediate or initial state.
After summation, the spin-flip processes cancel each other.
In this case $T_\sigma^{\rm dot,coh}(\omega) = |t_\sigma^{\rm dot}(\omega)|^2
= T^{\rm dot}_\sigma(\omega) = {\rm Re} \, [\Gamma_L \Gamma_R / 
(\omega - \epsilon + i0^+)^2]$.

{\it Interferometry with a single QD.}
To support the results of our intuitive picture, we analyze quantitatively 
AB interferometers which contain one QD.
The total transmission probability $T^{\rm tot}_\sigma(\omega)$ through the 
AB interferometer is the sum of three parts: $T^{\rm dot}_\sigma(\omega)$ and 
$T^{\rm ref}=|t^{\rm ref}|^2$ 
for the transmission through the dot and reference arm (the latter is 
independent of energy $\omega$ and spin $\sigma$), and the 
flux-dependent interference part $T^{\rm flux}_\sigma (\omega)$.
Two kinds of geometries have been considered, one using a two-terminal 
setup \cite{Yacobi95} and the other an open geometry \cite{Schuster97}.
The two geometries have in common that numerous channels (characterized by the 
energy $\omega$ and spin $\sigma$) are probed simultaneously, hence the 
interference signal is the sum of many contributions.
To achieve fully destructive interference one needs to adjust the amplitude 
of the reference arm such that 
$T^{\rm dot}_\sigma(\omega) = T^{\rm ref}$ for {\it all} contributing energies.

We relate the flux-dependent linear conductance to the dot Green's 
function for the two-terminal geometry.
To model the transmission through the reference arm we add to the Hamiltonian
a term $H_{\rm ref} = \sum_{kq\sigma} (\tilde t a^\dagger_{k\sigma R}
a_{q\sigma L} + {\rm H.c.})$ with 
$2\pi \tilde t \sqrt{N_L N_R} = |t^{\rm ref}| e^{i\varphi}$.
The AB flux $\Phi$ enters via $\varphi = 2\pi \Phi e/h$ (in a gauge that leaves
the tunnel Hamiltonian of the QD $\Phi$-independent).
The current from the right lead is given by the time derivative of the 
electron number, $I = e d \langle \hat n_R \rangle / dt = i(e/\hbar) 
\langle [ \hat H, \hat n_R ] \rangle$.
The latter expression yields Green's functions which involve Fermi operators
of the right lead.
Using the Keldysh technique we relate these to the dot Green's function.
After collecting all terms and using current conservation we find 
\cite{Koenig_unpub} the surprisingly simple relation for linear response and 
first order in $\Gamma$ and $t^{\rm ref}$ (i.e., higher harmonics in $\varphi$ 
are dropped)
\begin{eqnarray}
\label{T_2terminal}
   T^{\rm flux,a}_\sigma(\omega) = 2\sqrt{\Gamma_L \Gamma_R} \, |t^{\rm ref}|
	\cos \varphi \, {\rm Re} \, G^{\rm ret}_\sigma (\omega) \,  .
\end{eqnarray}

For the second kind of interferometer it was shown \cite{Gerland00} (under the 
condition that the open geometry ensures that the reference arm and the 
applied bias voltage do not affect the QD) that 
\begin{eqnarray}
\label{T_4terminal}
   T^{\rm flux,b}_\sigma(\omega) = 2\sqrt{\Gamma_L \Gamma_R} \, |t^{\rm ref}|
	\, {\rm Re} \, [ e^{-i\theta} G^{\rm ret}_\sigma (\omega)] 
\end{eqnarray}
with $\theta = \varphi +\Delta\theta$, where $\Delta\theta$ is determined by 
the specifics of the interferometer.

While Eqs.~(\ref{T_2terminal}) and (\ref{T_4terminal}) are almost self-evident 
in the noninteracting case, it was not a priori clear that they should hold
for interacting systems as well. 

According to Eq.~(\ref{T_2terminal}) the conductance is always 
extremal at $\varphi=0$. 
Such a ``phase locking'' does not take place in the open-geometry setup:  
the AB phase at which the transmission is extremal can be continuously varied 
by tuning the energy of the dot level via a gate electrode.

In the absence of interaction the flux-sensitive interference part 
for the latter geometry is
\begin{equation}
   T^{\rm flux, b}_\sigma(\omega) = 2 |t^{\rm ref}| {\rm Re} \left[ 
	e^{-i\theta}
	{\sqrt{\Gamma_L \Gamma_R} \over \omega - \epsilon + i\Gamma/2}
	\right] \, .
\end{equation}
At low temperature, $k_BT \ll \max \{ \Gamma, |\epsilon|\}$, we can 
adjust $t^{\rm ref}$ such that $T^{\rm dot}_\sigma(\omega) = T^{\rm ref}$ for 
{\it all} contributing energies (up to corrections of order  
$k_BT / \Gamma$ and $k_BT / |\epsilon|$) which yields \cite{comment_4terminal}
\begin{equation}
\label{open noninteracting result}
   {\partial I^{\rm tot} \over \partial V}\bigg|_{V=0} \propto 4 {e^2\over h}
	{\Gamma_L \Gamma_R \over \epsilon^2 + (\Gamma/2)^2} \left[ 1 -
	\cos \left(\theta - \theta_0\right) \right].
\end{equation}
Here $0 < \theta_0 = \arctan\left( \Gamma / 2\epsilon \right) < \pi$.
There is a value of the flux at which full destructive interference is 
achieved.
When $k_BT \gtrsim \min \{ \Gamma, |\epsilon|\}$, the matching of {\it all} 
the transmission amplitudes does not work, and full destructive interference 
is not achieved.

Let us now consider cotunneling (when $|\epsilon| \gg \Gamma, k_B T$ applies).
Expansion of Eq.~(\ref{open noninteracting result}) leads to
\begin{equation} 
\label{cotunneling noninteracting open}
   {\partial I^{\rm tot} \over \partial V}\bigg|_{V=0} \propto 4 {e^2\over h}
	{\Gamma_L \Gamma_R \over \epsilon^2} \left[ 
	1 - {\epsilon \over |\epsilon|} \cos \theta \right]
\end{equation}
showing that cotunneling in the noninteracting case is fully coherent.
In the interacting case we find
\begin{equation}
\label{cotunneling interacting open}
   {\partial I^{\rm tot} \over \partial V}\bigg|_{V=0} \propto 4 {e^2\over h}
	{\Gamma_L \Gamma_R \over \epsilon^2} \left[ 
	1 - {\epsilon \over |\epsilon|} {\cos \theta \over 1+f(\epsilon)} 
	\right] \, .
\end{equation}
For the two-terminal interferometer we obtain exactly the same as
Eqs.~(\ref{cotunneling noninteracting open}) and 
(\ref{cotunneling interacting open}) but with $\theta$ replaced by $\varphi$ 
and the proportionality replaced by the equality sign.

The factor $1/[1+f(\epsilon)]$ indicates an ``interaction-induced'' asymmetry
associated with spin-flip cotunneling, in accordance with 
Eq.~(\ref{cotunneling coherent}).
We, therefore, conclude that both kinds of interferometers discussed so far
are suitable to distinguish coherent from incoherent cotunneling through a QD.
Moreover, this asymmetry is an efficient and rather robust probe of the spin
configuration of the QD (whether it has total spin 0 or 1/2).
The same information can be retrieved by the Kondo effect, however, under 
more demanding experimental conditions.

In most experimental situations, the number of dot levels participating in the 
transport exceeds $1$. 
At a distance $|\epsilon|$ away from resonance, in the cotunneling regime,
there are $2N$ relevant levels with $N\sim 2|\epsilon|/\Delta+1$
(here $k_BT, \Gamma \ll |\epsilon| < E_C$, the mean level spacing is $\Delta$, 
and the factor $2$ represents spin degeneracy).
The number of levels within the range defined by temperature is $2M$ with
$M\sim 2k_B T/\Delta+1$.
The ratio of the number of coherent channels to the total number of 
transmission channels is $(N+1)/(N+4 M^2)$ in the valley where the electron
number on the QD is even and $N/(N+4 M^2)$ when it is odd \cite{Koenig_unpub}.
As a consequence the coherent contribution vanishes for 
$k_BT \gg \sqrt{|\epsilon| \Delta /8}$.
Furthermore, the asymmetry between adjacent valleys diminishes for 
$|\epsilon| \gg \Delta/2$.

What about first-order transport, which dominates for
$k_BT \gg \Gamma, |\epsilon|$? 
The energy spread of electrons going through the reference arm is $k_BT$, 
while the width of the resonance through the QD is $\Gamma$;
hence, the matching of {\it all} the transmission amplitudes to the reference
amplitude does not work, and full destructive interference is not achieved.
There is, however, at least partial coherence to lowest order in $\Gamma$.
This manifestly contrasts with the sequential-tunneling picture which
describes lowest-order transport as a sequence of incoherent tunneling 
processes.
Thus, it does not take into account the coherence of the transmitted and 
reference beam, although it produces the correct transmission probability 
through a QD in the absence of a reference arm.

{\it Interferometry with two QDs.}
The conceptual difficulty to address first-order transport in the above 
geometries is that the temperature has to be on the one hand large, yet, on 
the other hand, it has to be small to allow for a destructive interference of 
all energy components simultaneously.
To circumvent this problem, we consider a two-terminal AB interferometer with 
two QDs, one in each arm.
Then, fully destructive interference (in the absence of interaction) is 
feasible at high temperatures.
In related work, resonant tunneling (in the absence of interaction and flux)
\cite{Shahbazyan94} and cotunneling \cite{Akera93,Loss00} has been studied
in the same geometry \cite{Izumida97}.

Each dot is described by the Hamiltonian introduced above for a single QD.
We choose a completely symmetric geometry, and we assume 
$k_B T \gg \Gamma,|\epsilon_1|,|\epsilon_2|$ as well as
$\Gamma \gg |\epsilon_1 -\epsilon_2|$, where $\epsilon_{1,2}$ is the energy of 
the level in QD 1 and 2.
In this regime lowest-order transport dominates, and we can set 
$\epsilon = \epsilon_1 = \epsilon_2$.
To model the enclosed flux we attach a phase factor $e^{i\varphi/4}$ to the
tunnel matrix elements $t_{R, \rm QD 1}$ and $t_{L, \rm QD 2}$, and 
$e^{-i\varphi/4}$ to $t_{L, \rm QD 1}$ and $t_{R, \rm QD 2}$.
The system is equivalent to one QD with two levels (each of them spin 
degenerate) with $\varphi$-dependent tunnel matrix elements.
The total current is \cite{Meir92}
\begin{eqnarray}
   I^{\rm tot} = {ie \over 2h} && \int d \omega \, {\rm \bf {tr}} \left\{
	\left[ {\bf \Gamma}^L f_L - {\bf \Gamma}^R f_R \right] {\bf G}^>
\right. \nonumber \\ && \left.
	+ \left[ {\bf \Gamma}^L (1-f_L) - {\bf \Gamma}^R (1-f_R) \right] 
	{\bf G}^<
	\right\}
\end{eqnarray}
with ${\bf \Gamma}^{L} = {\Gamma \over 2} \left( \begin{array}{cc}
1 & e^{+ i\varphi/2} \\ e^{- i\varphi/2} & 1 \\ \end{array} \right)
\delta_{\sigma \sigma'}$ and 
${\bf \Gamma}^{R} = \left( {\bf \Gamma}^{L}\right)^*$.
The matrices account for the two QDs.
Expansion up to linear order in the transport voltage $V$ and in the 
intrinsic line width $\Gamma$ 
yields
\begin{eqnarray}
   {\partial I^{\rm tot} \over \partial V}\bigg|_{V=0} &=& 
	- {4\pi e^2 \over h} \Gamma \int d \omega \, \left\{
	f' (\omega) A_{11}(\omega) - {\sin(\varphi/2) \over \pi}
\right. \nonumber \\ &&\times \left. 
	\left[ f(\omega) \,
	{\partial G^{>}_{12}\over \partial (eV)}
	+ \left[ 1 - f(\omega) \right] 
	{\partial G^{<}_{12}\over \partial (eV)}\right]
	\right\}
\label{two dots first order}
\end{eqnarray}
with 
$G_{12}^>(\omega) = G_{12}^<(\omega) =  2\pi i P_2^1 \delta (\omega-\epsilon)$
and $A_{11}(\omega) = \delta (\omega-\epsilon)$ in the absence and 
$A_{11}(\omega) = \delta (\omega-\epsilon)/[1+f(\epsilon)]$ in the presence
of interaction.
The off-diagonal density-matrix elements 
$P_2^1 = \big\langle |2\rangle \langle 1| \big\rangle$ vanish in equilibrium, 
but they are present for finite bias voltages.
To determine them we use a real-time transport theory developed in 
Ref.~\cite{Koenig96,Koenig99} and solve a generalized master equation.
We find \cite{Koenig_unpub} at $V=0$ and in zeroth order in $\Gamma$ that
$\partial P^1_2 / \partial (eV) = - (i/ 2) f'(\epsilon) \sin (\varphi/2)$ in 
the absence and
$\partial P^1_2 / \partial (eV) = - (i/ 2) f'(\epsilon)/[1+f(\epsilon)]^3 
\sin (\varphi/2)$ in the presence of interaction.
As a consequence, in the absence of an AB flux, only equilibrium Green's 
functions enter Eq.~(\ref{two dots first order}).
In the presence of flux, however, it is crucial to first account for finite 
voltage {\it nonequilibrium Green's functions}, and take the zero-bias limit 
only at the end.    

We find for the non-interacting case
\begin{equation}
\label{two dots noninteracting}
   {\partial I^{\rm tot} \over \partial V}\bigg|_{V=0} = 
	2 \, {\partial I^{\rm dot} \over \partial V}\bigg|_{V=0}
	\times \left[ 1 - \sin^2(\varphi/2) \right]
\end{equation}
with $(\partial I^{\rm dot} / \partial V)|_{V=0} = - (\pi e^2 / h) 
\Gamma f' (\epsilon)$ being the conductance through a single QD.
At $|\sin (\varphi/2)| = 1$ the total current vanishes, indicating that 
lowest-order transport is fully coherent.
This completely contrasts the picture of incoherent sequential tunneling.
In the absence of interaction, however, the transport should be fully coherent.
For the simple limit $U=0$ we can rederive Eq.~(\ref{two dots noninteracting})
by using Eq.~(7) of Ref.~\cite{Meir92}, determining the dot Green's 
function by an equation-of-motion approach, and expanding the result up to 
first order in $\Gamma$.

In the presence of interaction we obtain
\begin{equation}
\label{two dots interacting}
   {\partial I^{\rm tot} \over \partial V}\bigg|_{V=0} = 
	2 \, {\partial I^{\rm dot} \over \partial V}\bigg|_{V=0} \times
	\left[ 1 - {\sin^2(\varphi/2) \over [1+f(\epsilon)]^2} \right]
\end{equation}
with $(\partial I^{\rm dot} / \partial V)|_{V=0} = - (\pi e^2 / h) \Gamma 
f' (\epsilon) / [1+f(\epsilon)]$.

We point out that the total conductance is always smaller than the sum of
the conductances through the QDs taken apart. 
The factor $1/[1+f(\epsilon)]^2$ yields an interaction-induced asymmetry in 
the ratio of coherent to total transport around a conductance peak.

{\it Conclusion.}
We have shown that interactions lead to an asymmetric 
suppression of destructive interference.
In second-order transport we related this explicitly to spin-flip processes 
which give rise to an incoherent contribution to the transmission probability.
Even in first-order transport, the transmission is at least partially coherent.
This statement is probably supported by the experiment of 
Yacoby {\it et al.} \cite{Yacobi95} in which AB oscillations were observed 
in that regime.

Our systematic analysis of how to describe the coherent components of physical 
observables in the presence of interaction (which is different from the way 
they can be accounted for in the absence of interaction), may pertain to other 
problems, such as the interpretation of the transmission phase through an AB 
interferometer \cite{Yacobi95,Schuster97}.

We acknowledge helpful discussions with B. Altshuler, N. Andrei, P. Coleman,
Y. Imry, H. Schoeller, and G. Sch\"on.
This work was supported by the Deutsche Forschungsgemeinschaft under grant 
KO 1987-1/1, the U.S.-Israel Binational Science Foundation, the Minerva 
Foundation, the ISF founded by the Israel Academy of Sciences and 
Humanities-Centers of Excellence Program and by the DIP foundation.


\begin{references}
\bibitem{renormalization}
The renormalization terms derived in second order, ``cotunneling at 
resonance'' \cite{Koenig_cotunneling}, are negligible in this regime.
\bibitem{comment_reg}
The regularization $+i0^+$ is put here by hand, but can be derived within a 
complete cotunneling theory \cite{Koenig_cotunneling}.
\bibitem{comment_4terminal}
For the open geometry only a fraction of the electrons passing through the 
interferometer will reach the drain.
We therefore use a proportionality sign (rather than an equality) for the 
current in this setup.
\end{references}
\end{document}